\documentclass{epl}
\usepackage{graphics,epsfig}
\begin{document}

\title{Temperature chaos in a replica symmetry broken spin glass model
- {\large A hierarchical model with temperature chaos -}}
\shorttitle{Chaos in a replica symmetry broken spin glass}
\shortauthor{M. Sasaki and O.~C.~Martin}
\author{M. Sasaki$^1$ and O.~C.~Martin$^{1,2}$}
\institute{
 $^{1}$ Laboratoire de Physique Th\'eorique et Mod\`eles Statistiques,
b\^at. 100, Universit\'e Paris-Sud, F--91405 Orsay, France.\\
 $^{2}$ Service de Physique de l'\'Etat Condens\'e\\
Orme des Merisiers --- CEA Saclay, 91191 Gif sur Yvette Cedex, France.\\
}

\date{\today}

\maketitle

\begin{abstract}
Temperature chaos is an extreme sensitivity of the equilibrium 
state to a change of temperature. It arises in several
disordered systems that are described by the so called scaling
theory of spin glasses, while it seems to be absent in
mean field models. We consider a model spin glass on a 
tree and show that although it has mean field behavior
with replica symmetry breaking, it manifestly has ``strong'' temperature chaos.
We also show why chaos appears 
only very slowly with system size.
\end{abstract}

\section{Introduction}
\label{sec:introduction}
The fragility of the equilibrium state to an infinitesimal change of temperature 
is commonly referred to as ``temperature chaos''~\cite{BrayMoore87}.
Having such fragility away from a phase transition point 
probably requires the system to be frustrated, 
but whether temperature chaos actually
arises in generic frustrated systems is still subject to controversy.
In the context of spin glasses~\cite{MezardParisi87b,Young98},
temperature chaos is shown to be present for models on
Migdal-Kadanoff lattices~\cite{BanavarBray87,NifleHilhorst92}.
Furthermore, the standard scaling 
theories~\cite{FisherHuse86,BrayMoore87,FisherHuse88}
suggest that this is a general property of glassy systems; in support of this, 
the Directed Polymer in a Random Medium~\cite{Halpin-HealyZhang86} (DPRM), 
which is well described by the (spin glass) scaling theories, 
is known to have temperature 
chaos~\cite{FisherHuse91,SalesYoshino02b}. 
On the other hand, the Random Energy Model~\cite{Derrida80} 
has no temperature chaos\cite{FranzNifle95}, and what happens in 
the Sherrington-Kirkpatrick (SK) mean-field model of 
spin glasses is still unclear. 
A replica calculation for the SK model suggests the presence of 
temperature chaos\cite{FranzNifle95}, but the numerics indicate 
no chaos or only very weak 
chaos~\cite{BilloireMarinari00,BilloireMarinari02,MuletPagnani01}.
Furthermore, a more recent calculation by
Rizzo~\cite{Rizzo01} shows that temperature chaos
is absent in perturbation theory 
about the critical temperature $T_{\rm c}$ to the orders computed. 
To clarify this question of 
temperature chaos in mean-field 
spin glasses, in this paper we study a specific mean-field-like model.
By determining the probability density
of overlaps for two real replicas at 
two different temperatures, we show that
this model has temperature chaos even though
it has a mean field behavior
with replica symmetry breaking. 
Our quantitative study also gives a coherent picture of chaos 
and suggests why chaos is so weak in general. 

\section{The model based on a tree}
In this paper, we focus on the model introduced in 
ref.~\cite{SasakiMartin02a}. 
It is very similar to the model of a polymer on a disordered Cayley tree 
studied by Derrida and Spohn\cite{DerridaSpohn88} 
(see also \cite{MajumdarKrapivsky00,DeanMajumdar01}); 
the differences are that values of both energy and entropy are assigned 
to each branch of the tree and each state thus has extensive entropy. 
It is also close to the Random-entropy 
Random-energy model~\cite{KrzakalaMartin02}; 
however, the energies and entropies are assigned 
hierarchically and the entropy 
is not introduced 
in an ad-hoc way. 

The model is constructed as follows. 
We consider a Cayley tree rooted at $O$. 
Each branch point $B$ (including $O$) creates $K$ branches which connect $B$ to 
its descendants. 
A tree with $L$ generations is obtained by repeating this 
procedure $L$ times. 
We regard the leaves (the bottom points) of the tree as the states of the system. 
A tree with $L$ generations 
has $K^L$ states. A random energy $\epsilon$ and a random entropy $\sigma$ 
are associated with every branch of the tree. The variables $\epsilon$ 
(respectively $\sigma$) are drawn independently from the same distribution 
$\rho_E(\epsilon)$ ($\rho_S(\sigma)$). The energy $E(B)$ 
(entropy $S(B)$) of a branch point $B$ is given 
by summing up the $\epsilon$'s ($\sigma$'s) of 
the branches which lie along the path connecting it to $O$. 
This means that the values of energy and entropy are correlated
hierarchically. The distance $d_{ij}$ of two states $i$ and $j$ is $d$ 
($d=0,1,\ldots,L$)
if their first common ancestor arises on the $d$-th layer counted from below. 
The overlap $q_{ij}$ is related to $d_{ij}$ by $q_{ij}=1-d_{ij}/L$, 
where $L$ is the number of generations of the tree. 

Note that our model is mapped onto Derrida and Spohn's 
model\cite{DerridaSpohn88} 
if we set $\rho_E^*(\epsilon')\equiv\int {\rm d}\epsilon {\rm d} \sigma 
\delta (\epsilon'-\epsilon+T\sigma)\rho_E(\epsilon)\rho_S(\sigma)$. Therefore, 
we can use the results in ref.~\cite{DerridaSpohn88} 
whenever we consider observables which depend on just one temperature.
(Of course the concern of this paper is almost exclusively
observables associated with two temperatures.) A consequence of this
mapping is that our model has a critical temperature $T_c$ below which 
it exhibits one 
step replica symmetry breaking (RSB): when $T<T_c$, the distribution 
of overlaps consists of two delta function peaks, one
at $0$ and one at $1$.

\section{Derivation of the overlap distribution with 
two different temperatures}
To study temperature chaos in this model, consider a given
realization of the quenched disorder (the random energies and
entropies); for that disorder, introduce two real replicas at
equilibrium, one at temperature $T$ and the other
at temperature $T'$, both temperatures being below $T_c$. Of interest is the 
probability distribution of the overlap of these two replicas. 
We want to know how this distribution depends on $L$ and on
the temperatures. We thus calculate 
the disorder averaged ``integrated probability'' 
to find the two replicas at a distance less or equal to $d$. 
This probability is explicitly defined as
\begin{eqnarray}
Y_{TT'}(L,d)&\equiv& \overline{\frac{1}{Z_T(L) Z_{T'}(L)}
\sum\nolimits_{ij/d_{ij}\le d}{\rm e}^{-X_T(i)-X_{T'}(j)}}.
\label{eqn:defYTT'}
\end{eqnarray}
In this expression, $\overline{\cdots}$ represents the disorder average, 
$X_{T}(i) \equiv E(i)/T-S(i)$ is the free-energy divided by $T$ of state $i$, 
and $Z_T(L)$ is the partition function at temperature $T$ for $L$ generations. 
Using an integral representation of $1/x$ for the two quantities
$Z_T(L)$ and $Z_T'(L)$, we can rewrite eq.~(\ref{eqn:defYTT'}) as 
\begin{eqnarray}
Y_{TT'}(L,d) &=& \int_{-\infty}^{\infty} {\rm d}u {\rm d}v F_{TT'}(L,d;u,v),
\label{eqn:relationYF}
\end{eqnarray}
\begin{equation}
F_{TT'}(L,d;u,v)\equiv\overline{\exp\left[-e^{-u}Z_T(L)-e^{-v}Z_{T'}(L)-u-v\right]
\sum\nolimits_{ij/d_{ij}\le d}{\rm e}^{-X_T(i)-X_{T'}(j)}}.
\end{equation}
We can use 
$\sum\nolimits_{ij/d_{ij}\le d}\exp[-X_T(i)-X_{T'}(j)]=\sum\nolimits_{B_d}
\exp[-X_T(B_d)-X_{T'}(B_d)]z_T(B_d)z_{T'}(B_d),
$
where
$B_d$ is a general branch point in the $d$-th layer (counted from below) and
$z_{T}(B)$ is the partition function at $T$ of the sub-tree rooted at 
a branch point $B$, in order to obtain
\begin{eqnarray}
F_{TT'}(L,d;u,v)&\equiv&
\overline{\exp\left[-e^{-u}Z_T(L)-e^{-v}Z_{T'}(L)-u-v\right]}\nonumber\\
&&\times \overline{\sum\nolimits_{B_d}\exp[-X_T(B_d)-X_{T'}(B_d)]
z_T(B_d)z_{T'}(B_d)}.
\label{eqn:iniEQforF}
\end{eqnarray}
From this equation, we find
\begin{equation}
F_{TT'}(d,d;u,v)=H_{TT'}(d;1,1;u,v),
\label{eqn:initialF}
\end{equation}
\begin{equation}
H_{TT'}(d;m,n;u,v)\equiv \overline{[e^{-u}z_T(B_d)]^m [e^{-v}z_{T'}(B_d)]^n 
\exp\left[-e^{-u}z_T(B_d)-e^{-v}z_{T'}(B_d)\right]}.
\label{eqn:defHn}
\end{equation}

We can calculate $H_{TT'}(d;m,n;u,v)$ (including $H_{TT'}(d;1,1;u,v)$ 
which appears in eq.~(\ref{eqn:initialF})) by the following recursion 
formulae. For $m=n=0$, it is not so difficult to find 
\begin{eqnarray}
H_{TT'}(0;0,0;u,v)&=&\exp[-e^{-u}-e^{-v}],
\label{eqn:Hrecursion1}\\
H_{TT'}(d+1;0,0;u,v)&=& {\tilde H_{TT'}}(d;0,0;u,v)^K, 
\label{eqn:Hrecursion2}
\end{eqnarray}
where for a general two variable function $g(u,v)$, we have defined
\begin{equation}
{\tilde g}(u,v)\equiv \int {\rm d}\epsilon {\rm d}\sigma 
\rho_E(\epsilon)\rho_S(\sigma) g(u+\epsilon/T-\sigma,v+\epsilon/T'-\sigma). 
\end{equation}
The recursion formula for general $m$ and $n$ is derived 
by applying the relation 
\begin{equation}
H_{TT'}(d;m,n;u,v)=\frac{\partial^m}{\partial u^m}
\frac{\partial^n}{\partial v^n} H_{TT'}(d;0,0;u,v)
\label{eqn:formulaH_n}
\end{equation}
to eqs.~(\ref{eqn:Hrecursion1}) and (\ref{eqn:Hrecursion2}). 
For example, the recursion formula for $H_{TT'}(d;1,0;u,v)$ is
\begin{eqnarray}
H_{TT'}(d+1;1,0;u,v)&=&\frac{\partial}{\partial u}
 {\tilde H_{TT'}}(d;0,0;u,v)^K\nonumber \\
&=&  K{\tilde H_{TT'}}(d;1,0;u,v) {\tilde H_{TT'}}(d;0,0;u,v)^{K-1}.
\end{eqnarray}

Finally, a method similar to the one used in ref.~\cite{SasakiMartin02a}
leads us to 
\begin{equation}
F_{TT'}(L+1,d;u,v)=K{\tilde F}_{TT'} (L,d;u,v) {\tilde H}_{TT'}(L;0,0;u,v)^{K-1}
\hspace{5mm}(L\ge d).
\label{eqn:finalresult}
\end{equation}

In summary, the disorder averaged distribution of distances
$Y_{TT'}(L,d)$ can be computed by the following procedure:
(i) Calculate $H_{TT'}(d;1,1;u,v)$ (=$F_{TT'}(d,d;u,v)$) 
by evaluating numerically 
the recursions which are derived by applying eq.~(\ref{eqn:formulaH_n}) 
to eqs.~(\ref{eqn:Hrecursion1}) and~(\ref{eqn:Hrecursion2}). 
(ii) Calculate $F_{TT'}(L,d;u,v)$ by using the recursion 
eq.~(\ref{eqn:finalresult}). 
(iii) Compute $Y_{TT'}(L,d)$ by estimating numerically the integral 
in eq.~(\ref{eqn:relationYF}).

\section{Temperature chaos}
\begin{figure}
\begin{center}
\epsfig{figure=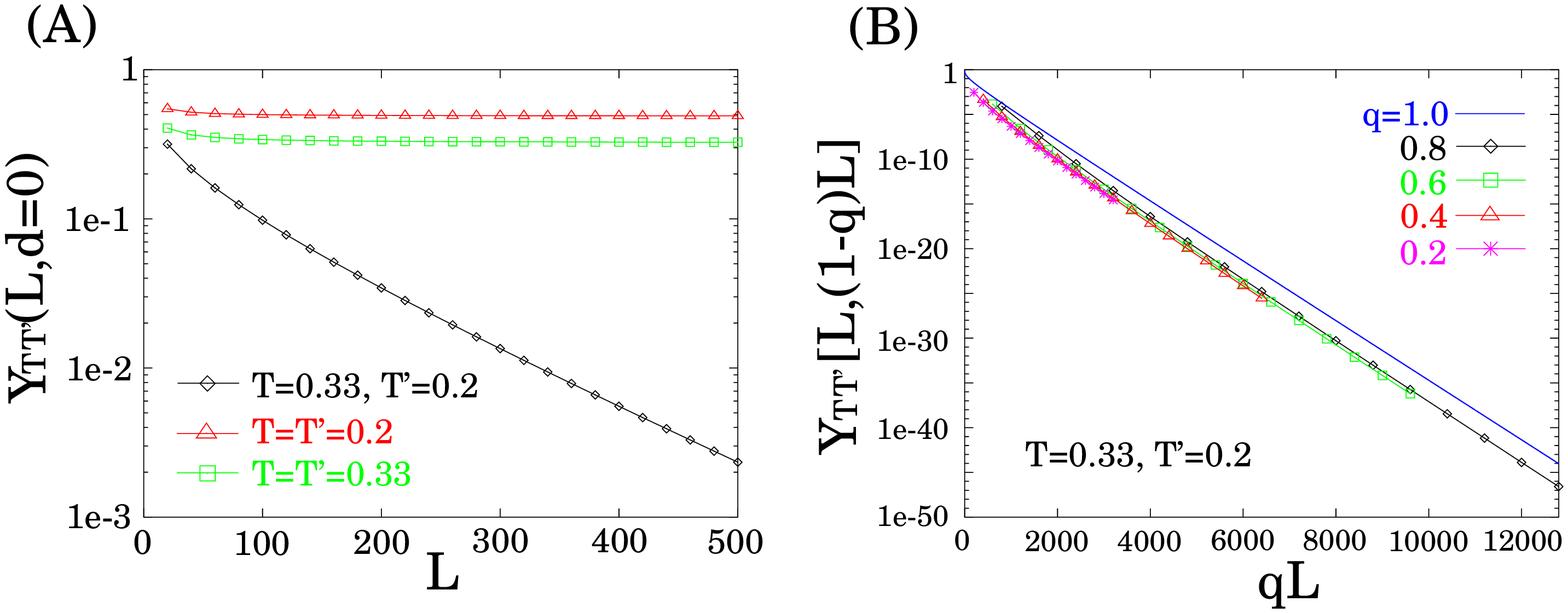,width=14cm}
\end{center}
\caption{(A) $Y_{TT'}(L,d=0)$ vs. $L$ 
for $(T,T')=(0.33,0.2)$, $(0.2,0.2)$ and $(0.33,0.33)$. 
The parameters for these data are $K=2$, $\rho_E(\epsilon)=0.25\delta(\epsilon)
+0.5\delta(\epsilon-1)+0.25\delta(\epsilon-2)$ and 
$\rho_S(\sigma)=0.5\delta(\sigma)+0.5\delta(\sigma-4)$. 
The critical temperature $T_{\rm c}$ is around $1.63$. 
(B) $Y_{TT'}(L,(1-q)L)$ for $(T,T')=(0.33,0.2)$ vs. 
$qL$. The data are taken for $q=0.2,0.4,0.6,0.8$ and $1.0$ with 
the same parameters as before. 
}
\label{Fig:chaos}
\end{figure}

To show that this model has temperature chaos, 
let us first measure
$
Y_{TT'}(L,d=0)=\overline{\sum_{i} P_{T}^{\rm eq}(i)P_{T'}^{\rm eq}(i)},
$
where $P_{T}^{\rm eq}(i)=\exp[-X_{T}(i)]/Z_{T}$. 
This is a generalization of $\overline{\sum_{i} \{P_{T}^{\rm eq}(i)\}^2}$ 
which has been studied in many systems like the SK model\cite{MezardParisi84} 
and the Random Energy Model~\cite{DerridaToulouse85}. 
The result is shown in Fig.~\ref{Fig:chaos} (A). 
We used $K=2$, 
$\rho_E(\epsilon)=0.25\delta(\epsilon)+0.5\delta(\epsilon-1)
+0.25\delta(\epsilon-2)$ 
and $\rho_S(\sigma)=0.5\delta(\sigma)+0.5\delta(\sigma-4)$
for those data. 
The critical temperature $T_{\rm c}$ is around $1.63$ by 
the mapping to the Derrida-Spohn model and using the
corresponding formula in~\cite{DerridaSpohn88}.
We see that $Y_{TT'}(L,d=0)$ 
decays exponentially for $T\ne T'$ while it 
converges to a non-zero value for $T=T'$. (More precisely,
a fit of the data at large $L$ gives
$Y_{TT'}(L,d=0) \approx A L^{-1/2} \exp (-B L)$.)
These results tell us that 
{\it the partition function below $T_{\rm c}$ is dominated by a few states, 
but these dominant states change with temperature}, i.e., there is 
temperature chaos. 
We have also checked that temperature chaos is absent in 
the model without entropy (no $\rho_S(\sigma)$);
this is in agreement with ref.~\cite{FranzNifle95} 
which shows that the GREM does not have temperature chaos. 
Ref.~\cite{FranzNifle95} also has shown that there {\it is} chaos 
against magnetic field in the GREM. But this result is not so surprising if one notices that 
the energy of state $i$ under field $H$ is $E(i)-HM(i)$ 
($M(i)$ is the magnetization of state $i$), 
and field plays the same role as temperature in our model. 

The quantity $Y_{TT'}(L,d=0)$ decays ``very fast'', in fact exponentially
with $L$, not as a stretched exponential or as a power of $L$.
This suggests that the overlap probability distribution itself
decays exponentially to zero for non-zero overlaps; to 
check this, we now study $Y_{TT'}(L,d)$.
Interestingly, it turns out that
$Y_{TT'}(L,d)$ satisfies the scaling law
\begin{equation}
Y_{TT'}(L,d)\approx {\hat Y}_{TT'}(L-d)
\label{eqn:scaling}
\end{equation}
for large $L$ and $d$. To show this, first rewrite eq.~(\ref{eqn:defYTT'}) as 
\begin{equation}
Y_{TT'}(L,d)=\overline{\frac{\sum_{B_d}\exp\bigl[-X_T(B_d)-\delta f_{T}(B_d)
 -X_{T'}(B_d)-\delta f_{T'}(B_d)\bigr] }{\sum_{B_d,B_d'}
\exp\bigl[-X_T(B_d)-\delta f_{T}(B_d)-X_{T'}(B_d')-\delta f_{T'}(B_d')\bigr]}},
\label{eqn:proofSCALE}
\end{equation}
where $\delta f_{T}(B_d)\equiv -\log[z_{T}(B_d)]+\overline{\log[z_{T}(B_d)]}$. 
Derrida and Spohn\cite{DerridaSpohn88} prove that 
$\delta f_{T}(B_d)$ has a limiting distribution that has 
a finite variance as $d\rightarrow \infty$. Furthermore, 
the statistics of $X_T(B_d)$ only depend on $L-d$. 
These facts lead us to the scaling law eq.~(\ref{eqn:scaling}). 

The validity of eq.~(\ref{eqn:scaling}) is confirmed 
in Fig.~\ref{Fig:chaos} (B) 
where $Y_{TT'}(L,(1-q)L)$ for $q=0.2,0.4,0.6,0.8$ and $1.0$ is plotted 
as a function of $qL$. Notice that $Y_{TT'}(L,(1-q)L)$ is the probability 
that the overlap of the two replicas is larger than $q$. 
The data, except those for $q=1$, satisfy the 
scaling very well (note that $Y_{TT'}(L,0)$ is calculated by 
eq.~(\ref{eqn:proofSCALE}) with $\delta f_{T}=\delta f_{T'}=0$). 
Furthermore, we see that the slopes 
for $Y_{TT'}(L,0)$ and for the scaling function are the same. 
This means that the presence of $\delta f_{T}$ in 
eq.~(\ref{eqn:proofSCALE}) does not change the slope because 
the variance of $\delta f_{T}$ is finite. Hereafter we regard 
the inverse of the exponent in this exponential decay 
as the chaos length $\ell(T,T')$ of the model. 

This analysis shows that $\int_{q}^{1} {\rm d}q' P_{TT'}(q')$ 
decays as $\exp [-q L/\ell(T,T')]$ if $q\ne 0$, 
meaning that $P_{TT'}(q)$ also decays 
(up to power corrections) exponentially. 
This property corresponds to ``strong'' chaos in 
any reasonable classification of chaos. 
To obtain some insight into the origin of the strong chaos,
let us focus on $Y_{TT'}(L,d=0)$ which is the sum
over all $K^L$ states of $\exp[-\{F_T(i)-F_{\rm eq}(T)\}/T - 
\{F_{T'}(i)-F_{\rm eq}(T')\}/T']$. (In this expression,
$F_T(i)$ is the free-energy of state $i$ at temperature $T$ 
and $F_{\rm eq}(T)$ is the equilibrium free-energy.) 
Now let us assume that among these $K^L$ states it is enough 
to consider just those that dominate the partition function at 
some temperature $T''$. Since they are dominant states, 
the energy, the entropy and the 
free-energy of these states are the same as the equilibrium ones 
at $T''$. Therefore, at any temperature $T_{\rm m}$ the free-energy 
$F_{T_{\rm m}}(i)$ of these states are $E_{\rm eq}(T'')-T_{\rm m}S_{\rm eq}(T'')$
 $(=F_{\rm eq}(T'')-S_{\rm eq}(T'')(T_{\rm m}-T''))$. 
On the other hand, the Taylor expansion of $F_{\rm eq}(T_{\rm m})$ around $T''$ 
leads us to $F_{\rm eq}(T_{\rm m})=F_{\rm eq}(T'')-S_{\rm eq}(T'')(T_{\rm m}-T'')
-\frac{1}{2T''}C(T'')(T_{\rm m}-T'')^2+{\cal O}((T_{\rm m}-T'')^3)$,
where $C$ is the heat capacity. 
By using this for $T_{\rm m}=T$ or $T'$, we find that 
the contribution to $Y_{TT'}(L,d=0)$ for such a state
is $\exp[-\{(T-T'')^2/T + (T'-T'')^2/T'\} C(T'')/(2T'')]$. 
For $\Delta T\equiv T-T'\ll 1$, this is maximized at $T''=\frac{T+T'}{2}$ 
and we obtain 
\begin{equation}
\label{eq_naive_ell}
Y_{TT'}(L,d=0)\approx\exp[-\Delta T^2 C(T)/(4T^2)].
\end{equation}
In our model, $C$ grows linearly with $L$, leading to an exponential 
decay of $Y_{TT'}(L,d=0)$ with $L$. 
On the contrary, the specific heat in the low temperature phase 
is zero in the REM\cite{Derrida80} and in our model 
without entropy\cite{DerridaSpohn88} and thus there is no chaos 
in these systems. 

Interestingly, this computation is only qualitatively correct
and eq.~\ref{eq_naive_ell} does {\it not}
give the exact overlap length. The reason is that
we have relied on typical contributions to $Y_{TT'}$
while in fact it is dominated by rare events:
a tiny fraction of the samples where the same state is dominant 
at $T$ and $T'$ determine the disorder averaged probability $Y_{TT'}$. 
To calculate the true $\ell(T,T')$, 
we have to take into account such rare events; to do so, we
first study how fluctuations grow with $L$ at a given temperature. 

\section{Scaling of the entropy fluctuations}
\begin{figure}
\twofigures[width=7cm]{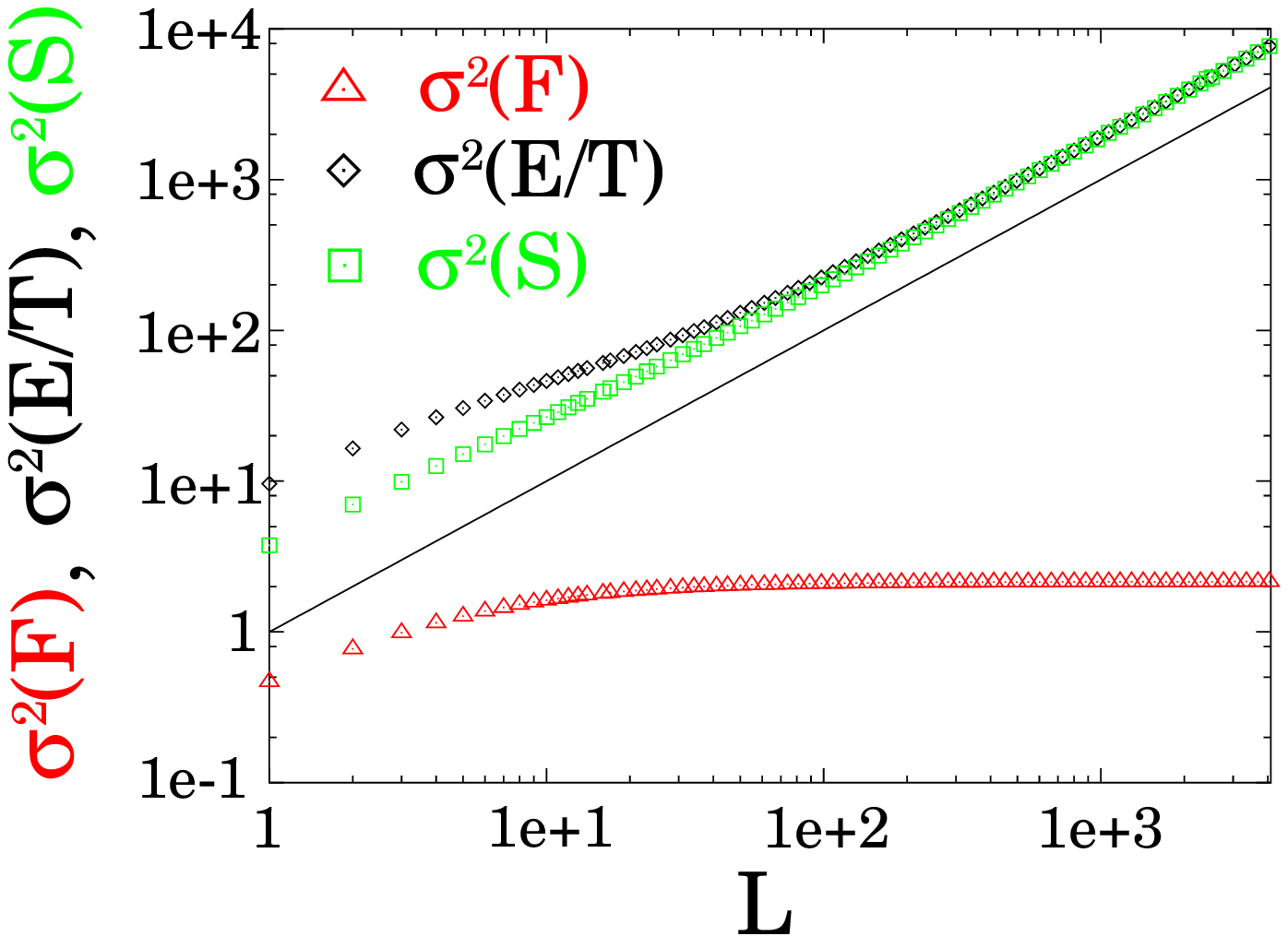}{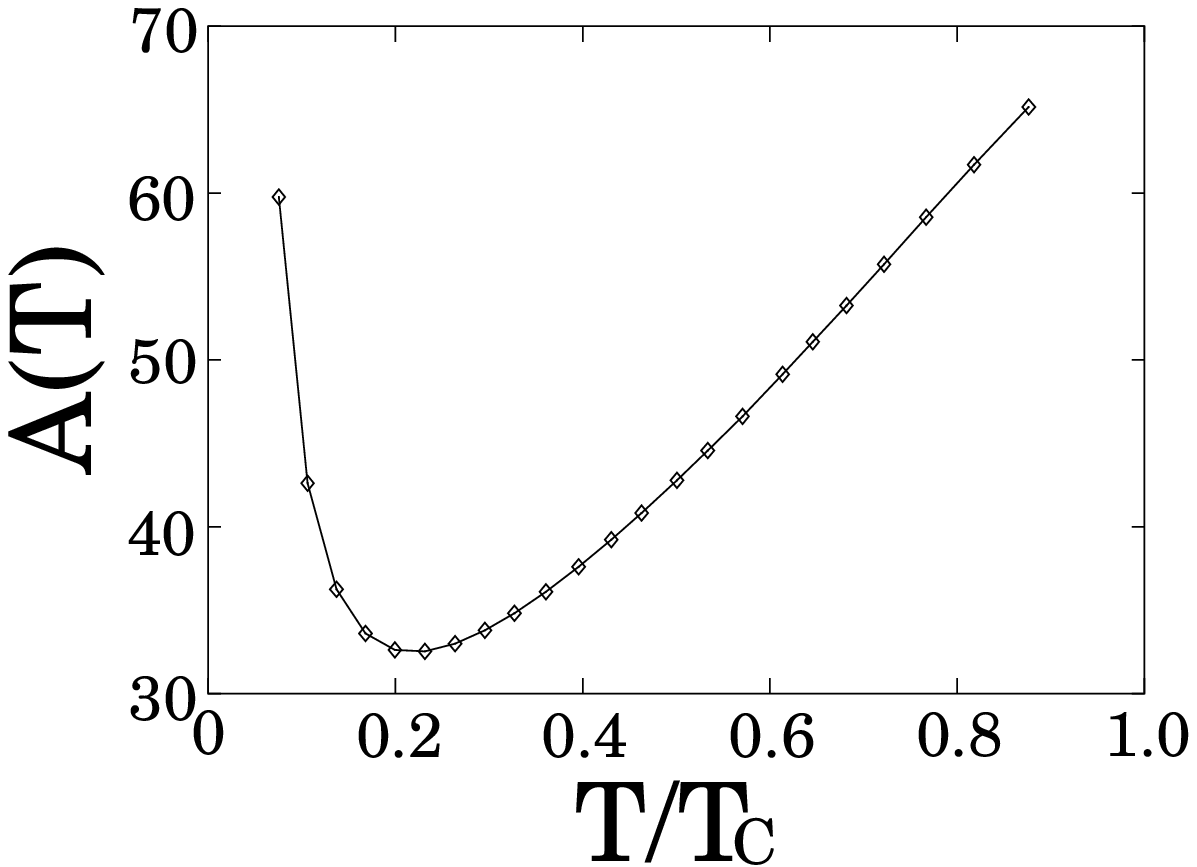}
\caption{Fluctuations of energy, entropy and free-energy at $T=0.2$ vs. 
generation $L$. The parameters are the same as in Fig.~\ref{Fig:chaos}. 
A function linear in $L$ is drawn to guide the eye. 
}
\label{Fig:Fluctuation}
\caption{A(T) vs. $T/T_{\rm c}$ using the 
same parameters as Fig.~\ref{Fig:chaos}. 
$A(T)$ is defined in eq.~(\ref{eqn:lastR2}). 
}
\label{Fig:chaoslength}
\end{figure}

In Fig.~\ref{Fig:Fluctuation}, we show the fluctuations 
of entropy, energy and free-energy which are defined as 
\begin{equation}
\sigma^2_T({\cal O})=\overline{\langle {\cal O}^2 \rangle_T} 
-\left\{\overline{\langle {\cal O}\rangle_T}\right\}^2,
\label{eqn:defsigma}
\end{equation}
where ${\cal O}$ is quantity associated with {\it each} state, i.e., 
energy, free-energy and entropy, and 
\begin{equation}
\langle{\cal O}\rangle_T \equiv \frac{\sum_{i}{\cal O}(i)\exp[-X_T(i)]}{Z_T(L)}.
\end{equation}
These quantities were calculated by recursion formulae 
similar to the ones for $Y_{TT'}$. We clearly see that 
$\sigma^2_T(S)$ and $\sigma^2_T(E)$ grow linearly with $L$, while $\sigma^2_T(F)$ 
converges to a finite value. These results show that 
there are a few states which have almost the same lowest free-energy, 
but whose entropies are very different from one-another. 
Therefore, the relative order of these dominant states can change
by a small change of the temperature, i.e., 
the free-energy levels can cross, and these kinds of crossings generate 
temperature chaos in this model. Note that this mechanism 
of temperature chaos was first proposed in the scaling 
theories\cite{FisherHuse86,BrayMoore87,FisherHuse88} 
and its validity is also confirmed 
in other systems\cite{SalesYoshino02b,KrzakalaMartin02}. 
It is also worth noticing that 
$\Delta S(i,T)\equiv S(i)-\overline{\langle S \rangle_T}$ 
and $\Delta E(i,T)$ are strongly correlated for the dominant states so that 
$\Delta S(i,T)\approx \Delta E(i,T)/T$ because of 
the relation $\Delta F(i,T)=\Delta E(i,T)-T\Delta S(i,T)$. 
This is the reason why $\sigma^2_T(S)$ and $\sigma^2_T(E/T)$ are 
almost the same in Fig.~\ref{Fig:Fluctuation}.

The same results hold for state-to-state fluctuations defined as 
$\hat\sigma_T^2({\cal O})\equiv \overline{\langle{\cal O}^2 \rangle_T}
-\overline{\langle {\cal O}\rangle_T^2}$. 
First, $\hat\sigma_T^2(F)$ stays $O(1)$ 
since $\hat\sigma_T^2(F)\le\sigma_T^2(F)$. 
Second, $\hat\sigma_T^2(E)$ grows linearly with $L$ because 
$\hat\sigma_T^2(E)$ is proportional to the heat capacity. 
From these two results, 
$\hat\sigma_T^2(S)\approx\hat\sigma_T^2(E/T)\propto L $. 

\section{Consequences for the overlap length}
Let us estimate $Y_{TT'}(L,0)$ to 
calculate $\ell(T,T')^{-1}$
(recall that $Y_{TT'}(L,0) \sim\exp[-L/\ell(T,T')]$). We denote the state 
with the lowest free-energy at $T$ 
by $D_T$. If $F_{T'}(D_T)=E(D_T)-T'S(D_T)$ happens to be smaller than 
$\overline{\langle F \rangle_{T'}}$, the dominant state at $T'$ is still 
$D_T$ ($D_{T'}=D_T$) 
so that $Y_{TT'}(L,0)$ 
for that sample is of order 1. 
Therefore, 
\begin{eqnarray}
Y_{TT'}(L,0)&\approx& {\rm Prob}(F_{T'}(D_T)\le \langle F \rangle_{T'})\nonumber \\
&=& {\rm Prob}\left((T-T')\Delta E(D_T,T)\le T[\overline{\langle F \rangle_{T'}}-
\overline{\langle F \rangle_{T}}-(T-T')\overline{\langle S \rangle_{T}}]
\right),
\end{eqnarray}
where we have used $\Delta S(i,T)\approx \Delta E(i,T)/T$ 
to go from the first line to the second. Now assume that 
the distribution of $\Delta E(D_T,T)$ is Gaussian; 
this seems to be plausible since $\sigma_{T}(E)$ is linear in $L$, 
as if there was an underlying
central limit theorem process. Then we obtain 
\begin{equation}
Y_{TT'}(L,0)\sim \{L/\ell(T,T')\}^{-\frac12}\exp[-L/\ell(T,T')],
\end{equation}
\begin{equation}
\ell(T,T')=\frac{2\sigma_{T}^2(E)L(T-T')^2}
{T^2\left[\overline{\langle F \rangle_{T'}}-
\overline{\langle F \rangle_{T}}-(T-T')\overline{\langle S \rangle_{T}}\right]^2}.
\label{eqn:exponent}
\end{equation}
The accuracy of eq.~(\ref{eqn:exponent}) was checked by comparing 
$\ell(T,T')$ estimated from $Y_{TT'}(L,0)$ and 
from eq.~(\ref{eqn:exponent}). The result was very satisfactory, 
i.e., the former is $131.3$ and the latter $131.8$ 
when the parameters are those used in Fig.~\ref{Fig:chaos}. 
We also found similarly good accuracy for the other sets of $(T,T')$
we tested. From eq.~(\ref{eqn:exponent}), we find 
\begin{equation}
\ell(T,T+\Delta T)\approx A(T)\left(\frac{\Delta T}{T}\right)^{-2}
\hspace{5mm}(\Delta T\ll 1), 
\label{eqn:lastR1}
\end{equation}
\begin{equation}
A(T)=8\sigma_{T}^2(E) C(T)^{-2}T^{-2} L,
\label{eqn:lastR2}
\end{equation}
where again $C(T)$ is the heat capacity
\footnote{Rigorously speaking, eq.~(\ref{eqn:lastR2}) is valid when 
$\overline{\langle F \rangle_{T}}/L= -k_{\rm B} T \overline{\log Z(T)}/L$. 
This relation is justifiable in the low temperature phase where 
only a few states dominate thermodynamics of the system, and 
we have checked this numerically.}. 
It should be noted that
the chaos exponent $\zeta$ defined
via $\ell(T,T+\Delta T) \approx (\Delta T)^{-{1/\zeta}}$ 
is correctly given 
by the droplet theory\cite{FisherHuse88,SalesYoshino02b} 
which predicts $\zeta=\frac{d_{\rm s}-2\theta}{2}$ 
if $\hat\sigma_{T}^2(S)\propto L^{d_{\rm s}}$ and 
$\hat\sigma_{T}^2(F)\propto L^{2\theta}$. 
(Indeed, Fig.~\ref{Fig:Fluctuation} shows 
$d_{\rm s}=1$ and $\theta=0$ in this model). 
Figure~\ref{Fig:chaoslength} shows $A(T)$ of the model. We find that 
$A(T)$ has a minimum around $T\approx 0.2$ for which the value is about $33$. 
This tells us that temperature chaos emerges only at large scales, 
e.g., when temperature is changed by $10\%$ ($\frac{\Delta T}{T}=0.1$), 
the chaos length is at least $3300$. But note that 
eqs.~(\ref{eqn:lastR1}) and (\ref{eqn:lastR2}) give us chaos {\it volume}
if we consider the case $d\ne 1$ since $L$ is volume in this model. 
(Remember that energy and entropy are proportional to $L$.) 
Therefore, the minimum chaos {\it length} for $\frac{\Delta T}{T}=0.1$ 
is not so large for $d=3$, i.e., $3300^{1/3}\approx 15$. 

\section{Conclusions}
We have studied a GREM-like system with extensive entropy;
it has strong temperature chaos, 
$P_{TT'}(q)$ decaying as $\exp[-qL/\ell(T,T')]$ if $T\ne T'$. 
Entropy fluctuations from valley to valley
are the central ingredients for temperature chaos, 
as predicted by the scaling (droplet) 
theory~\cite{FisherHuse86,BrayMoore87,FisherHuse88}. 
Note that the overlap length $l(T,T')$ 
is proportional to $C(T)^{-1}(T-T')^{-2}$ (see eqs.~(\ref{eq_naive_ell}) 
and~(\ref{eqn:lastR2})) and that $C(T)$ is typically small. 
If $C(T)$ controls the decay of overlap probability in more general disordered 
systems also, then it is no surprise that 
temperature chaos is difficult to detect in simulations. 
Finally rejuvenation and memory effects observed in off-equilibrium 
dynamics\cite{JonasonVincent98,NordbladSvedlindh98} are 
naturally interpreted by this model because 
it has both temperature chaos and a hierarchical structure. 
Consider for example the case where temperature is changed as 
$T\rightarrow T-\Delta T \rightarrow T$. A strong {\it rejuvenated} 
relaxation will be observed at $T-\Delta T$ due to temperature chaos, while 
memory will emerge when the temperature is returned to $T$ 
because of the hierarchical structure.

\section{Acknowledgments}

We thank J.-P. Bouchaud, F. Krzakala, H. Yoshino, and 
especially
M. M\'ezard for helpful discussions.
M. S. acknowledges a fellowship from the French Ministry of research. 
The LPTMS is an Unit\'e de Recherche
de l'Universit\'e Paris~XI associ\'ee au CNRS.

\bibliographystyle{prsty}
\enlargethispage{30pt}
\bibliography{references}

\end{document}